# Fog Computing in IoT Aided Smart Grid Transition- Requirements, Prospects, Status Quos and Challenges

Md. Muzakkir Hussain[1], Mohammad Saad Alam[2], M.M. Sufyan Beg[1]
[1]Department of Computer Engineering, ZHCET, AMU
`md.muzakkir@zhcet.ac.in`
`mmsbeg@cs.berkeley.edu`
[2]Department of Electrical Engineering, ZHCET, AMU
`saad.alam@zhcet.ac.in`

**Abstract:** Due to unfolded developments in both the IT sectors viz. Intelligent Transportation and Information Technology contemporary Smart Grid (SG) systems are leveraged with smart devices and entities. Such infrastructures when bestowed with Internet of Things (IoT) and sensor network make universe of objects active and online. The traditional cloud deployment succumbs to meet the analytics and computational exigencies decentralized, dynamic cum resource-time critical SG ecosystems. This paper synoptically inspects to what extent the cloud computing utilities can satisfy the mission critical requirements of SG ecosystems and which subdomains and services call for fog based computing archetypes. The objective of this work is to comprehend the applicability of fog computing algorithms to interplay with the core centered cloud computing support, thus enabling to come up with a new breed of real-time and latency free SG services. The work also highlights the opportunities brought by fog based SG deployments. Correspondingly, we also highlight the challenges and research thrusts elucidated towards viability of fog computing for successful SG Transition.
**Keywords:** *Smart Grid (SG), Internet of Things (IoT), Fog computing, Cloud computing.*

## 1 Introduction

The dawn of the Smart Grid (SG) has acquired the global consensus because of the intrinsic shortcomings associated with the century old hierarchical power grid [1]. The shortcomings arise due to primitive generation methodologies, irregular generation-consumption profiles, underutilization of infrastructure resources, security breaches and global climatic concerns etc [2], [3]. The SG aims to resolve the voltage sags, overloads, blackouts and brownouts caused due to erratic nature of power consumption and demand, rise in complexity of power system networks triggered by varying modalities such as distributed generation, dynamic energy management (DEM), electric vehicles, micro-nano grid etc [4]. The SG congregates electrical and communication network elements to enable noble and bidirectional energy cum data flows across the whole infrastructure [3], [5], [6].

The future SG operation is envisioned to be more data reliant than the electrical power [7]–[9]. The self-governing Communicate Interact and Operate (CIO) protocols established by geo-distributed intelligent network nodes in a data aware SG architecture will ensure hassle free back and forth power delivery to and from the SG stakeholders. The entities such as SCADA systems, smart meters from advanced metering infrastructures (AMI), roadside units (RSU) and on board units (OBU) from electrically run transportation telematics and miscellaneous sensors etc dispersed across multi-dimensional networks such as Home Area Networks (HAN), Metropolitan Area Networks (MAN), and Transport Oriented Cities (TOC) etc, form the prime source of data generation and consumption in a typical SG infrastructure [11-13].

A smart SG architecture is equipped with facilities for real-time, duplex communication between producers and consumers, having software modules to control and manage the power usage at both ends [4], [5]. Meanwhile, the current landscape of IoT penetration into smart grid architectures, alters the data generation and consumption profiles [10]. The components of SG now called to be "things" will produce galactic mass of data while in execution, thus urging for robust storage and compute framework that can process and serve the service requests of SG stakeholders. Since the major percentage of IoT endpoints in a SG are primitive i.e. the deployment of required compute and storage resources is not guaranteed everytime, an external agent should undertake the computation and analytics tasks. The storage and processing loads in a typical SG is swarmed up from billions of static and mobile nodes spanning over vast geographical domain. Such heterogeneity in the data architecture of a smart grid envisaged the use of advanced computing technology to overcome technical challenges at different levels of computation and processing. Rather than relying on the master-slave computation model as in legacy systems, the current notion is to get switched to data center level analytics operating under client-server paradigm [1], [11], [12].

The objective of reaching a consensus on where to install the compute and storage resources prevails as an open question for the academia, industries, R&Ds and legislative bodies [13], [14]. The cloud computing emerged to be promising technology to support SG because of its ability to provide convenient and on-demand, anytime, anywhere network access to its shared computing resources. provisioned and released with minimal management effort or service provider interaction [6], [15], [16]. The cloud service also frees the IoT devices from battery-draining processing tasks by availing unlimited pay-per use resources through virtualization [15]. However, the varying modalities of services facilitated by cloud computing paradigms perish to meet the mission critical requirements of SG [1], [2]. The existing Cloud Computing paradigm is perishing to welcome its proponents because of its failure in building common and multipurpose platform that can provide feasible solutions to the mission critical requirements of SG in IoT space.

The main driving force for the emergence of FOG (From cOre to edGe) computing model (FC) is to abridge the computational hysteresis prevalent in cloud computing where the elastic cloud resources are extended to the end-points of the network [17], [18]. In the context of SG, the fog computing can be defined as an architectural setup for federated as well as distributed processing where application specific logic is embedded not only in remote clouds or edge systems, but also across the intermediary infrastructure components such as portable devices, gateways, smart meters, RSUs and OBUs at electric vehicular networks, wireless sensors at smart homes and micro-nano grids, and miscellaneous IoT devices [19]. Brute force analysis of the requirement specification plans (RSPs) of smart grid architectures reveals that the mere of fact of migrating and executing everything to the mega datacentres creates the prime unfeasibility concerns. Thus the notion is to develop platforms for IoT aware smart grid architectures where significant proportion of compute and storage activities will be offloaded to geodistributed nodes named as fog nodes. A robust fog topology allows dynamic augmentation of associated fog nodes, thus rejuvenating the elasticity and scalability profiles of mission critical infrastructures. This paper outlines the fog computing paradigm and examines its primacy over the cloud computing counterpart that became ubiquitous in fulfilling the computational and analytics needs of a reliable, robust, resilient and sustainable SG.

Having the assumption that the power system community is not in a position to design its own Internet infrastructure or install its own computing platforms from scratch, and hence must work with generally accepted standards and commercially successful hardware and software platforms, this paper inspects to what extent the cloud computing utilities can satisfy the mission critical requirements of SG ecosystems and which subdomains and services call for fog based computing archetypes. Obviously, rigorous research and investment effort are required to rollout the century old legacy power grid with a robust and reliable SG and the objective of this research is to assess how far the current cloud computing platforms are equipped to meet the needs of this transition and which attributes of current computing industry needs to be augmented or revised using fog approaches. **The contribution of the paper can be outlined as**

**1.** Analyzed the existing cloud infrastructures to see if they respond to SG computing requirements and in which aspects they need fog reforms.
**2.** Demystified the fact of how fog (FC) paradigms can serve as an ally to cloud platforms and assess how far such a noble mix of both computation models will successfully satisfy the high assurance and mission critical computing needs of SG.
**3.** Outlined the significant research & development (R&D) opportunities generated along with the adoption challenges encountered towards realizing the commercial viability and optimal implementation of any fog model proposed for SG.

## 2    Cloud-Fog Computing for Computationally Robust Smart Grid- A Synoptic Overview

The SG entirely modernizes the energy generation, transmission, distribution and consumption landscapes of power system. These four key SG activities are executed across Home Area Networks (HAN), Neighborhood Area Networks (NAN) and Wide Area Networks (WAN). The HAN is the layer nearest to ground, that consists of smart devices, home appliances (including washing machines, televisions, air conditioners, refrigerators and ovens), electrical vehicles, as well as renewable energy sources (such as solar panels). The HANs are built within residential units, in industrial plants and in commercial buildings and connect electrical appliances with smart meters and they manage the consumers' on-demand power requirements those sub-domains. The NAN(s) connect smart metering devices across multiple HANs and supports communication between distribution substations and field electrical devices for power distribution systems. The WAN being the uppermost layer serves as a communication backbone between network gateways or aggregation points. The data from NAN data collector nodes are aggregated in WAN layers to facilitate proper synchrony among transmission systems, bulk generation systems, renewable energy sources and control center.

Let us consider Smart Homes as a representative SG use case to show how large-scale computing must play a key role in the smart power grid. The IoT technologies will result high adoption of SG powered home appliances such as smart heating/refrigeration system, smart infotainment services, home security sub-system and smart emergency response devices like lighting control, fire alarms, temperature monitoring devices etc [20], [21]. The integration of IoT services will enhance transparency in the infrastructure by tracking the activities of the smart city residents. These smart homes will be leveraged with a range of AMI and monitoring devices ensuring fine-tuning of consumption patterns with power tariffs and load surges on the power grid [21].

The IoT sensors and actuators dispersed across the smart HANs sense the data associated to that environment and offload to the controlling module operated by householders [22]. Often the smart homes are also equipped with utilities that assimilate the surveillance data in order to predict the future occurrence of events, thereby preparing the householders to behave according to the contingencies. For instance, a geyser might be on when power is cheap but the water is allowed to cool when hot water is unlikely to be needed. An air conditioning, washing machine system or even smart TVs might time themselves to match use patterns, power costs, and overall grid state [23]. In fact, there seems to be a one-one mapping between a SCADA system and a smart home and having the realization former directly reaches into the smart homes. The smart homes or customers may also form communities through Neighbor Area Network (NAN) to have access to shared resources and response intelligently to the community-based cooperative energy consumption schemes [24]. The scope of this paper is to summarize the status of cloud based SG solutions coupled with application domains and benefits of fog computing for successful SG rollout. In this section we analyse the current computing requirements of SG in six key metrics namely Decentralization, Scalability, Consistency, Latency, Privacy & Security, and Availability and Reliability. Correspondingly, we also assess the status of cloud based solutions and discuss the suitability of Fog based solutions while satisfying those requirements.

**2.1  Decentralization:**

The data generation and consumption nodes in SG architectures are sparsely distributed. In addition to a centralized control, the sensor and actuator nodes deployed across the smart home like sub-systems demand geo-distributed intelligence. The information transparency in distribution domains may need to be extended from mere SCADA systems to a scale that ensures national level visibility. The platform as a service (PaaS) resources can be hired to install the SCADA systems, a significant portion of which can be shared across power providers and distributed generation sources. The data center based SG analytics will maintain transparency at varying levels of granularity. The web applications running in software as a service (SaaS) mode will allow authenticated web access to track the status of IoT endpoints. Collaboration of cloud computing technologies leveraged with data analytics modules into electrified transportation (Vehicle to Grid mode) use-cases will ensure robustness and resiliency in penetration of EV fleet of any size. The current centralized cloud deployments fail to capture the context of the decentralized SG computing. Since the protocols for visibility delineations are defined by third party cloud vendors, discrepancies may arise duo to biased favors. As an illustrative example, let us take the V2G scenario where the EVs participate in energy market acting as prosumers (both producer and consumer). The EVs will discharge its power into the energy market following an incentive policy. Since the V2G integration and recommendation policy is done via aggregators, there may be the case that the cloud market policy is more inclined to aggregator's payoff and the EV customers may be deprived of appropriate benefits. Moreover, in data center based centralized control policies, the degree of transparency enforcements is still in the hand of business giants. Under such circumstances there may be a slender abuse of decentralization.

The fog computing platforms will potentially realize both context and situational awareness through the intelligence incorporated in the geo-distributed Fog Computing Nodes (FCN).The power generating bodies will utilize the synchrophasor data from decentralized SCADA components such as Phasor Measurement Units (PMUs) and phasor data concentrators (PDCs) processed in FCNs in order to have context aware operational visibility of power grid dynamics and prepare themselves to respond intelligently to any disturbances that can lead to sudden blackouts or brownouts. For the same V2G example, the EV end user may from communities and rely on local computations performed in associated FCN. The EVs will be dependent not only on the "hidden hands of energy market" but on the outputs its own recommendation algorithms.

**2.2  Scalability**

The IoT deployments unifies multitude of such controllable entities into a single pulpit all demanding computational scalability as an indispensable necessity. These smart homes may behave as independent candidates for intelligent SCADA control. The EVs may also be coordinated via home energy management systems (HEMS), micro-grid integration, indicating computational scalability concerns. The IoT equipped autonomous vehicles must tap to the dynamic data generated by SG utilities, viz. state of charge (SOC) and load predictions, tariff structures and miscellaneous attributes that ensure grid stability. All such instances create potential thrusts for scalability enforcements and demand a paradigm shift into new computational paradigms.

The data center based computational transformation has resulted in the displacement of traditional SCADA based state estimation paradigms. The traditional SCADA like computational platforms derelict because of the scalability problem associated with SG components when deployed on a large scale. The cloud system are bestowed with large numbers of lightweight, inexpensive servers thus becoming irreversibly dominant to support scalable analytical solutions from large numbers of sensors, actuators, customers and other SG stakeholder's data. A single cloud computing data center might have storage and computing capabilities tens or hundreds of times greater than all of the world's supercomputing facilities combined. This incentivizes the SG community to migrate high performance computing (HPC) applications at these data centers. Over time, cloud PaaS services leaves counter-incentive prospects for industry investment in faster HPC systems. Moreover the horizontal as well as vertical extendibility support provided by cloud platforms enables the SG utilities to expeditiously respond to any market or regulatory changes and to the state of the art products and services.

Talking about scalable-consistency preserving SG requirements, the current cloud vendors seem unilaterally focused towards scalability challenge, are deploying massive but weak storage and processing configurations that often "embrace inconsistency". For high assurance applications they seem to be annoyed by scalability consistency dilemma. The business behavior is more dedicated to the motto "serve more customers" rather than "serve more critical customers". The cloud utilities are tuned with stale data to respond to user's vectored requests and the later required to deal with this. Fog models leverage massive scalability support from multi-hypervisor virtualized systems with bandwidth optimized paths. Fog architectures are equipped with Fog Nodes (FNs) that rely on three key technology enablers for scalable and efficient virtualization of the key resource classes, thus enabling full manifestation of SG distributed architecture. These are:

i) *Compute:* In order to virtualize both the computing and I/O resources, proper selection of compute resources such as hypervisors is necessary.

ii) *Network:* For efficiently managing the network resources listed above, viable and robust Network Virtualization Infrastructures such as Software Defined Networking (SDN) and Network Functions Virtualization (NFV) techniques should be employed.

iii) *Storage:* It defines a Virtual File System (VFS) and a Virtual Block and/or Object Store.

**2.3     Consistency:**

The demand for consistency preserving decision making analytics is being emerged as the need for the hour to carry up mission critical requirements of SG utilities. For SG, consistency is a broad term associated with ACID (Atomicity, Consistency, Isolation and Durability) guarantees, support for state machine replication, virtual synchrony, and deployments having only limited count of node failures. To understand the consistency need of SG use-cases, consider an EV fleet where multiple vehicles are concurrently communicating to SCADA system, even if the formers are under distributed control and the communication may also across multiple networks, they all should receive the right control instructions. Minute deviation from the communication synchrony may lead to catastrophic deformation in the fleet dynamics. Ubiquitous network access provided by cloud platforms guarantee the access of everyone to the cloud and procure interpretability between the SG systems. A common communication platform that is offered by the cloud provides data flow in SG and avoids the use of multiple middleware software and interfaces to access the data by SG systems. Data consistency is also supplied with standardization of data formats on one central platform. However, the often cloud giants employ the embrace of the Consistency Availability and Partitioning (CAP) [3] to justify their consistency-scalability tradeoffs. The SG applications feed into a service spectrum supporting wide degree of shared access thus poor scaling of consistency is not feasible. Though the CAP theorem can also be formalized under weaker assumptions, but the clouds make stronger assumptions in practice and cite such folk logic for offering weak consistency guarantees, although they consider strict consistency assurances for their own needs. With introduction of fog layer, inconsistency issues in cloud architecture can be solved to great extent. Operating systems and applications may support fog computing by using device level and application level architectures respectively. Community server will definitely play crucial role in achieving offline computation. Architectures can be alternatives in some cases, while complimentary to each other in some other cases. Depending on requirements and context further modifications can be done.

### 2.4    Real-Time Analytics:

Based on the data latency hierarchy requirements, the SG applications can be grouped into three categories with loosely defined boundaries. Installation of transmission infrastructures, power delivery road maps etc. come under group A type applications with relaxed timing requirements. For such application current computing configurations guarantee service level agreements (SLA) because of informal service constraints. Group B applications are those that needs super paced communication and transport channels such as circulating the smart meter data to regulate SCADA control. Such applications are tolerable to only unit to a few tens of microsecond's delays due to caused due to node failures or connectivity disruptions. Such SG solutions demand real-time response even in the presence of failures. The third class C includes mission critical SG applications that require high assurance, stringent privacy and security enforcements, robust access control and consistent behavior across the nodes of action. The applications acting on real-time data may produce glitches if exposed stale data. The SCADA system employed in a modern data driven SG are so timed that it can malfunction when operated over ubiquitous TCP/IP protocols and cannot sustain the out-of-the-box TCP flow control latencies.

In order to have a detailed, real-time information on the state of devices, the current SG infrastructures are planned with deployment of a large number of sensors and smart meters in various points of the grid is planned. For real-time monitoring of SG applications the cloud data centers obtain two-way communication and relay data from wireless sensor networks (WSN), disseminates it for proactive diagnosis and timely response any erroneous context that can lead to transient faults or blackouts in worst case. On the transmission and distribution side, the cloud utilities can be employed collect real-time information that provides wide-area situational awareness (WASA) of power grid status while from the consumer's perspective; real-time estimates of anticipated usage through advanced metering infrastructure (AMI) enable demand response (DR) controls thus enhancing energy utilization efficiency. To have an uninterrupted SG dynamics, any computing paradigm must be adept to the timing constraints and even if some servers' fiascos occur, the utilities should heal themselves with just graceful degradation in latency services. The contemporary cloud platforms are well matched to group A type applications described in section II. Though current cloud systems do support group B like services requiring real-time responses, but the response time can be disrupted by transient Internet congestion events, or even a single server failure, whereas it still lacks the technology for hosting applications that come under group C. Fog enables data analytics at the network edge and can support time-sensitive functions for SG like cyber-physical systems. This is essential for developing not only stable control systems but also for the tactile Internet vision to enable embedded AI applications with millisecond response requirements. Such advantages in turn enable new services and business models, and may help broaden revenues, reduce cost thereby accelerating IoT aided SG rollouts.

### 2.5    Privacy and Security:

The SG utilities when leveraged with IoT endpoints needs robust protective mechanisms that ensures restricted and entrusted access to critical data. The SG confidential data might be of interests to the criminals or manipulating entities seeking edge to energy trade, thus exposing the system to cyber-attacks. The denial-of-service (DoS) or distributed denial-of-service (DDoS) attacks on AMI or vehicular data may cause severe vulnerabilities such as bandwidth drainage, excessive CPU utilization, irregular memory surges, and halting the client's or host's operations etc [25]. The designed computing architectures should be leveraged with robust aggregation algorithms that can guarantee the privacy and data anonymization of respective stakeholders and motivate consumer's participation [26]. Other security aspects that need to be addressed properly in smart home, micro-grid and electrified transportation use cases are data outage, threat detection, and cyber-physical attacks etc [15].

In context of SG *applications security here* is to mean the safety and stability of the power grid, rather than protection against malice which come under the privacy umbrella. The malignancy of casual justification of CAP theorem is manifested in the current position of smart grid cloud security. In current designs all data finds its way into cloud storages comprised of huge number servers and storage elements having peculiar horizontal and vertical elasticity. Unfortunately, the existing cloud based security and privacy enforcements are precisely erratic, and many a times the cloud operators became evil if they are intended to do so. In a competitive shared cloud SG environments the worry is that the rivals may spy on property data, leading to cyber-physical war (CPW) in extreme cases. Such platforms though always guarantee that the data center will be on and the applications will keep running, but when analyzed at finer depths they are still far from ideal for discrete data items and individual computations.

Gartner claimed that the cloud platforms are fraught with security risks and suggests SG like customer must put rigorous questions and specifications before the cloud service providers. They should also consider a guaranteed security assessment from a neutral third party prior to making any commitment. The woeful protection services of current cloud deployments often stimulate the cloud vendors to recapitulate their security management folks to "not be evil". Rigorous efforts are on headway across the power system and transportation communities to come up with SG cloud utilities and platforms leveraged with robust protective contrivances where the stakeholders could entrust the storage of sensitive and critical data even under concurrent share and access architectures [21].

## 2.6 Availability Reliability and High Assurance Computing:

A subset of SG applications need to have hardware and software solutions that are "everytime on". Any interruption in computing services may cause increased costs and many a times it may lead to loss of consumer confidence. Because cloud computing by its nature relies primarily on the internet connectivity, the SG utility vendors interested in starting or expanding their business strategies with current cloud platforms must have rigorous IT consultations for showing them ways to schedule network resources such as bandwidth levels which will suffice their mobility and availability constraints [6], [14]. While conducting technical operations in the remote datacenters the control of SG stakeholders over the infrastructure will be seized. Thus the manageability is not as robust as when they are used to with SCADA workstations and other controllers.

As according to the resource protocols employed in cloud computing the user's physical control over the outsourced data is relinquished permanently, hidden data breaches from untrusted cloud vendor(s) may jeopardize SG data confidentiality, data integrity, and data availability. Switching from traditional power grid to SG subsystems when introduced to an IoT universe endangers the energy sector with substantial risk factors, an issue that need to get fixed in its inception. The penetration of autonomous EVs for instance into the modern road transport creates huge bowls of multi-dimensional data sets, an asset which if mishandled, may befool the execution of whole systems. In fact, the data generated due to Cloud-IoT integrated transportation telematics and advanced metering infrastructures (AMI) can prove to be harmful to its stakeholders, specifically for privacy and security [28]. Thus, it's an earnest need for the stakeholders to be assured with stringent protection protocols and be inert from the vulnerabilities. Such scenario necessitates incorporating robust risk analysis procedures that will evaluate and quantify the computational and business risks that persist in such critical infrastructures. Selection followed by implementation of proper risk analysis paradigms is itself a full-fledged realm to dwell on. Risks perceived to be minor in inception phase, later elicits tougher public concerns. Though the "pay-for-usage" protocols of cloud computing business models are efficient in satisfying the bulky analytics and computational tasks, the bliss transforms into worries when the applications demand null-latency services and when the data stream chokes the bandwidth **restricted communication buses.** The aforesaid discussions yet cloud computing, as we've shown, lacks key properties that power control and similar smart grid functionality will need. These include security, consistency, real-time assurances, ways to protect the privacy of sensitive data, and other needs.

## 3    Adoption Challenges and Future Research Opportunities

In this section we inspect the opportunities brought by fog based SG deployments. Correspondingly, we also highlight the challenges and research thrusts generated towards viability of fog computing for successful SG Transition.

### 3.1    *Opportunities*

#### 3.1.1.    Compute and Connectivity Convergence

Fog computing enables the base station (BS) of the cellular network specifically F-RANs to deploy both ad-hoc and dedicated Fog Computing Nodes (FCN) for leveraging the SG data consumers with the compute and temporary storage services. The FCN embedding in BS(s) allows the later to process computation and communication workload thereby bringing benefit in terms of performance for diverse applications. Convergence of computation and communication resources will enhance the performance of numerous SG applications such as:

**i)** Actionable real time intelligence in micro-grid systems such as wind farm, solar panels etc.
**ii)** Real-time outage management in generation-distribution subsystems.
**iii)** Dynamically optimizing the wind farm operation through   forecasting at finer granularities (once every 5 minutes or less).

The compute and connectivity convergence further consents the underlying SG networks architectures to utilize the context information of communication to perform desired classification mechanisms on the huge amount of generated data by IoT endpoints.

#### 3.1.2.    Flexibility in Application development

Software and application developers can play productive roles in the success of fog computing implementations. They can design and develop ground breaking applications and services, which can take the benefits of context information available in fog radio network (F-RANs) particularly at the fog gateways. Application developers can design their applications more flexible for the fog environment by using the application programming interfaces, open standards, and software development kits (SDKs). The applications available for the cloud environment are mainly available on the server. However, in case of fog paradigms, the application should be able to migrate from the mobile device to the fog server/nodes for the execution. Therefore, the application developers have to redesign the applications by separating the delay sensitive compute-intensive parts from the delay tolerant parts of the application. This creates a room for the application developers to take the benefits from the specific demands of fog to take a share in the success of the emerging paradigm.

#### 3.1.3.    New Revenue Stream for Service Providers

The SG software providers can deploy new services for mobile applications and enterprises to leveraging them to improve their productivity. The context-awareness of local network can introduce new service categories and enhance the features of services offered to SG stakeholders. The introduction of new services can add new revenue streams for the fog service providers as well SG vendors. The deployment of applications and services in the HANs will not only enhance the end user application performance but also shrinks the volume of signaling traffic to the core network, thus minimizing the operational cost of the SG service providers.

### 3.1.4. Opportunities to Network Equipment Vendors

Network elements, such as fog gateways and access points, need more computational power to facilitate the end users with on-demand resources. The network equipment vendors have more demands for manufacturing the network elements as the service providers will have to replace the existing base stations and access points with computationally rich resources. Fog platforms will not only open new avenues for application developers and network service providers but also for the network equipment vendors to introduce their products with new features and capabilities. The addition of new features and capabilities will create a new revenue stream for the SG software vendors.

## *3.2 Challenges*

### 3.2.1. Compatibility, Scalability and Complexity:

The contemporary IoT based SG applications are diverse in terms of reliability, interoperability, scalability, and security. Location, configuration, and served functionalities of diverse FNs are key issues. Selecting the best IoT entities and fog components to compose an optimal application workflow while meeting non-functional requirements such as security, network latency, and QoS create nascent research thrusts across the computational domain. Since numerous IoT vendors and manufacturers are involved in developing heterogeneous sensors and smart devices, it is increasingly complicated to reach the consensus of how to select optimal components customized to SG hardware configurations and personalized requirements. There may be SG applications that require specific hardware and protocols and might even operate under high security constrained environments. An efficient orchestration framework can cater such disruptions in functionalities and requirements and successfully manage large and dynamic workflows. One key issue for the fog orchestrator is to determine whether the architectural elements like cloud resources, sensors, and FNs are capable of provisioning any complex services correctly and efficiently [30]. It should automatically predict, detect, and resolve issues pertaining to scalability bottlenecks that could arise from increased application scale in SG domain.

### 3.2.2. Security:

Security is among the prime issues for a typical data driven SG platform, as the transportation utility vendors may agonize for the repercussions if the privacy of entrusted cloud data is compromised. Due to the dynamic nature of a transportation and SG infrastructures, it becomes nearly infeasible to create coherent cross-cloud trust relationships. Further, existence of complex relationships and dependencies among varying range of stakeholders in contemporary smart cities hinder the compatibility and cost effectiveness of data clouds employed in Transport Oriented Cities (TOCs). Global security standards are essential to cope up with such privacy and flexibility concerns. Decisions regarding selective migration of information hosted in private clouds, FNs and vehicular cloudlets to the public storage space require rigor research. Robust and fine grained authorization protocols and access grants should be defined to ensure multiple accesses to federated cloud and federated fog repositories. In the typical IoT aware SG environment, multiple sensors, smart nano chips, and communication devices each deployed within different geographic locations are integrated to establish the overall communication. In such circumstances, the FNs are particularly prone to attack vectors, especially in the context of network enabled (e.g. SDN) IoT systems, where the attack vectors can include agent generated sabotage of network infrastructure, malicious programs provoking leakage of data streams, or even unauthorized access to physical devices.

*Table. 1: Challenges and Future research Opportunities in Fog Security*

| PRIVACY AND SECURITY CHALLENGES | RESEARCH OPPORTUNITIES |
|---|---|
| Malicious or Malfunctioning FCNs- Such nodes are potential sources of data breach | Design/adaptation of low-cost end-to-end security scheme(s) |
| Malicious insider attack – The attacker steals the end users private key and illegitimately accesses his data. | In order to identify any malicious insider, the consumer's behavior may be mined using usage mining to deploy decoy technology. The decoy method returns massive garbage that is illusioned as user's data to the intruder. Optimal placing of decoy in FCN networks and dynamically designing on-demand decoy information are future opportunities. |
| Mutual authentication among dynamic fog node and end-users | Since majority of user devices are mobile having random routing profile and FCNs often join and leave the network, as in VFC, mutual authentication of front end devices, consumer endpoints- and FCN is challenging. In case any user misbehave is detected by the controlling authority, how to keep the anonymity of the users and to trace the consumers with their true identity is an open research problem in fog settings. |
| Fog forensics- A technique that provides the digital evidence by reconstructing the past fog computing events. | Although the cross-border issue is less significant as compared to cloud computing due to distributed nature of fog computing, the fog forensics still requires international legislation and jurisdictions and application level logging. Thus, it is still an important task to overcome cross-border legislation challenges in fog computing. |
| Authentication and Key Agreement in Fog radio access networks (F-RANs)- Analogous to C-RANs the F-RAN is vulnerable to network attacks, such as replay attack, man-in-the-middle attack, and DoS attack. | • How to achieve scalable, authentication and billing in the context of F-RANs.<br>• How to design an authentication and key agreement protocol for F-RANs through SDN and NFV technologies?<br>• How to model the replay attack, man-in-the-middle attack, and DoS attack in F-RANs? |

In SG networks having millions of IoT endpoints, effective and accurate evaluation of threats and security compromises is crucial. When the workflows dynamically evolve and adapt at runtime, enforcing holistic security and risk assessment standards becomes tedious and challenging. Algorithms that can dynamically evaluate the security of SG application instances will become increasingly critical for secure data placement and processing.

Since in IoT aides SG like use-cases, a specific service is composed of multiple sensors, computer chips and devices, their deployment in varying different geographic locations result in increased attack vector of involved objects. The FCNs are vulnerable to range of attacks, a summarized description of which is given in Table 1. Examples of attack vector may be human-caused sabotage of network infrastructure, malicious programs provoking data leakage, or even physical access to devices. Holistic security and risk assessment procedures are needed to effectively and dynamically evaluate the security and measure risks, as evaluating the security of dynamic IoT based application orchestration will become increasingly critical for secure data placement and processing. The IoT integrated devices for fog support such as switches, routers and base stations etc, if are brought to be used as publicly accessible computing edge nodes, the risk associated by public and private vendors that own these devices as well as those that will employ these devices will need revised articulation. Also, the intended objective of such devices, e.g. an internet router for handling network traffic, cannot

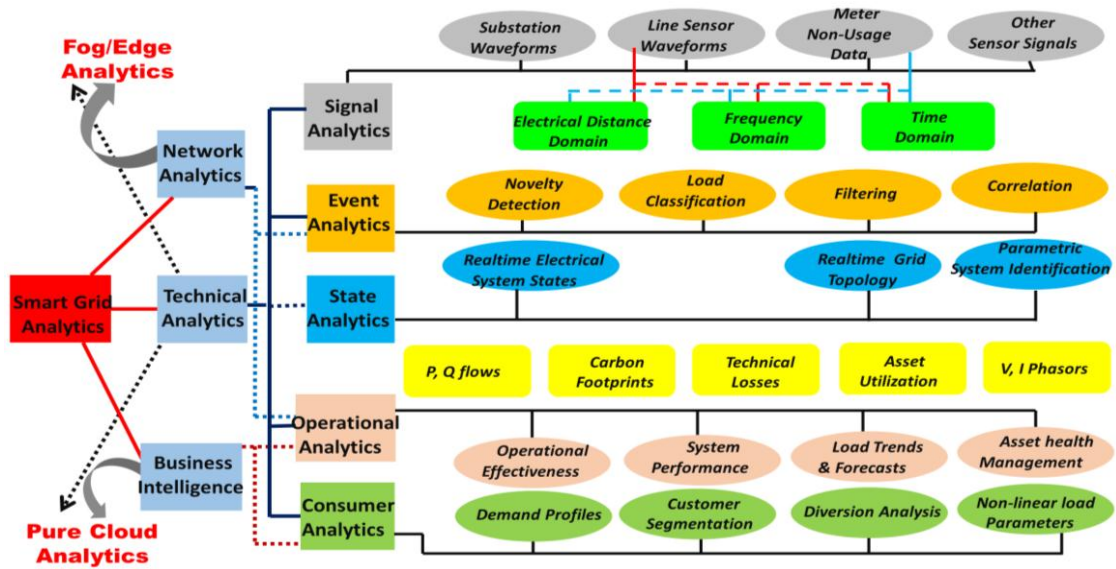

*Fig. 1: Landscape of Big Data Analytics that can be supported over Fog based Smart Grid Architectures*

### 3.2.3. Performance and Reliability:

Being numerous fog utility offerings available on the market with varying pricing schemes, decisions on selecting the one commercially optimal to SG entities needs to be standardized. A budding informatics thrust is to evaluate the complexity and financial viability of fog service deployments in price diverse environments. The infrastructure assembles multiple cloud and fog genres into a common platform, thus uncertainties in pricing models is obvious. The stakeholders if are aware of the future service tariffs and incentives, will allow them to ponder for the optimum. The incentives of Vehicle to Fog (V2F) like use-cases is primarily dedicated to promote development of intelligent vehicular services and offer a range anxiety free drive to the naïve EV users. The SG infrastructure assembles the distributed cloud and fog platforms to co-work with each other for smooth and reliable operation of its entities. However, maintaining an optimal balance in the distribution of data, control and computation among the dedicated cloud-cloudlets-FNs decides the performance of the whole system. Commercial realization of the notion of cloud of things (CoT) and fog of things (FoT) from billions of sensors and low power devices in a sensor network and connectivity with the data centers demand reliable and permanent sources of energy. Efficient fabrication techniques can enable the sensors to generate onsite power from renewables and environment. For V2F like infrastructures that employ dynamic and adhoc fog nodes, intermittent vehicular networking will hamper the service quality and non-functional QoS. The mesh created by seamless communication among cloud-cloudlet utilities will create galactic volumes of information to flow across the interfaces and data centers, thus uncertain network & communication failure will adversely affect the execution of backend SG infrastructure. Intelligent controllers and gateways coupled with mobile networking paradigms can manage the connectivity control of distributed and networked cloud and fog resources in SH based TOC cyber infrastructures. **3.2.4. Data Driven Analytics and Optimization:**

A typical SG framework congregates the diverse transportation entities into a clique like structure through CoT and FoT paradigm and enables a bidirectional flow of energy and data among the stakeholders in order to facilitate the assets optimization. The major data sources for a data driven SG include but not limited to:

**i)** SG data aggregation nodes such as Supervisory control and data acquisition (SCADA) system and associated components viz. master terminal unit (MTU), remote terminal unit (RTU), programmable logic controller (PLC) etc.

**ii)** AMI metering and sensing devices.

**iii)** ITS objects such as EV OBU, RSU, traffic sensors and actuators, GPS devices etc.

**iv)** Web data for recommender systems, crowdsourcing, feedback modules.

Furthermore, the domain of IoT applications in SG is extended to numerous geographically distributed devices that produce multidimensional, high-volume dynamic data streams requiring a noble mix of real-time analytics and data aggregation. The fog orchestration module should employ efficient data-driven optimization and planning algorithms for reliable data management across complex IoT aware smart grid endpoints. Figure 1 shows a landscape of possible Big Data Analytics (BDA) solutions to be adopted for SG, by employing fog assisted cloud frameworks as supporting processing framework. While designing SG applications adhered to FC and making proper trade of such applications across different layers in the fog environment the developers should employ robust optimization procedures that stabilizes the schema definitions, mappings, all overlapping, interconnection between layers (if any). In order to reduce data transmission latencies data processing activities and the database services may be pipelined. Rather than frequent triggering of Move-Data actions, use of multiple data- locality principles (e.g. temporal, spatial etc.) and efficient caching techniques can distribute or reschedule the computation tasks of FNs near the sensors thereby improving the delays. The data relevant attributes related to QoS parameters such as the data-generation rate or data-compression ratio can be customized to adapt to the desired degree of performance and assigned resources to strike a balance between data quality and specified response-time targets.

## 4      Conclusion

In this work, we analyzed the current state of Cloud Computing services in fulfilling mission critical computing demands of SG, and side by side we identified the requirements that could be fulfilled, if carried through fog computing. The work comprehends the opportunities and prospects of fog based SG analytics. Finally, the significant adoption challenges encountered towards fog deployment are outlined along with the future research avenues that will prove to be productive while going for fog based SG computing.

### References

[1]  Y. Saleem, N. Crespi, M. H. Rehmani, and R. Copeland, "Internet of Things-aided Smart Grid: Technologies, Architectures, Applications, Prototypes, and Future Research Directions," pp. 1–30.

[2]  K. Birman, L. Ganesh, and R. van Renesse, "Running Smart Grid Control Software on Cloud Computing Architectures 1 . Introduction : The Evolving Power Grid," *Work. Comput. Needs Next Gener. Electr. Grid*, pp. 1–28, 2010.

[3]  Gungor, V. Cagri Sahin, Dilan Kocak, Taskin Ergut, Salih Buccella, Concettina Cecati, Carlo Hancke, Gerhard P."A Survey on smart grid potential applications and communication requirements", IEEE Transactions on Industrial Informatics, vol. 9, issue 1, pp. 28-42

[4]  D. S. Markovic, D. Zivkovic, I. Branovic, R. Popovic, and ..., "Smart power grid and cloud computing," ... *Sustain. Energy ...*, vol. 24, pp. 566–577, 2013.

[5]  J. Gao, Y. Xiao, J. Liu, W. Liang, and C. L. P. Chen, "A survey of communication/networking in Smart Grids," *Futur. Gener. Comput. Syst.*, vol. 28, no. 2, pp. 391–404, 2012.

[6]  A. R. Al-Ali and R. Aburukba, "Role of Internet of Things in the Smart Grid Technology," *J. Comput. Commun. Technol. J. Comput. Commun.*, vol. 3, no. 3, pp. 229–233, 2015.

[7]  C. C. Chan, D. Tu, and L. Jian, "Smart charging of electric vehicles – integration of energy and

information," *IET Electr. Syst. Transp.*, vol. 4, no. 4, pp. 89–96, 2014.
[8] W. June, "Big Data Analytics to Support the Smart Grid," no. June, 2017.
[9] Z. Lv *et al.*, "Next-Generation Big Data Analytics : State of the," vol. 13, no. 4, pp. 1891–1899, 2017.
[10] D. T. Hoang, P. Wang, D. Niyato, and E. Hossain, "Charging and discharging of plug-in electric vehicles (PEVs) in vehicle-to-grid (V2G) systems: A cyber insurance-based model," *IEEE Access*, vol. 5, pp. 732–754, 2017.
[11] M. H. Rehmani, M. Erol Kantarci, A. Rachedi, M. Radenkovic, and M. Reisslein, "IEEE Access Special Section Editorial Smart Grids: a Hub of Interdisciplinary Research," *IEEE Access*, vol. 3, pp. 3114–3118, 2015.
[12] S. Xu, Y. Qian, and R. Q. Hu, "On Reliability of Smart Grid Neighborhood Area Networks," *IEEE Access*, vol. 3, pp. 2352–2365, 2015.
[13] L. Yu, T. Jiang, and Y. Zou, "Fog-Assisted Operational Cost Reduction for Cloud Data Centers," *IEEE Access*, vol. XX, no. XX, pp. 1–8, 2017.
[14] O. Osanaiye, S. Chen, Z. Yan, R. Lu, K. K. R. Choo, and M. Dlodlo, "From Cloud to Fog Computing: A Review and a Conceptual Live VM Migration Framework," *IEEE Access*, vol. 5, pp. 8284–8300, 2017.
[15] S. Bera, S. Misra, and J. J. P. C. Rodrigues, "Cloud Computing Applications for Smart Grid: A Survey," *IEEE Trans. Parallel Distrib. Syst.*, vol. 26, no. 5, pp. 1477–1494, 2015.
[16] Cisco Systems, "Fog Computing and the Internet of Things: Extend the Cloud to Where the Things Are," *Www.Cisco.Com*, p. 6, 2016.
[17] S. Sarkar, S. Chatterjee, and S. Misra, "Assessment of the Suitability of Fog Computing in the Context of Internet of Things," *IEEE Trans. Cloud Comput.*, vol. PP, no. 99, pp. 1–1, 2015.
[18] S. Misra and S. Sarkar, "Theoretical modelling of fog computing: a green computing paradigm to support IoT applications," *IET Networks*, vol. 5, no. 2, pp. 23–29, 2016.
[19] R. Emilia, R. Emilia, P. G. V. Naranjo, M. Shojafar, and L. Vaca-cardenas, "Big Data Over SmartGrid -A Fog Computing Perspective Big Data Over SmartGrid - A Fog Computing Perspective," no. November, 2016.
[20] A. Zanella, N. Bui, A. Castellani, L. Vangelista, and M. Zorzi, "Internet of Things for Smart Cities," *IEEE Internet Things J*, vol. 1, no. 1, pp. 22–32, 2014.
[21] R. Ullah, Y. Faheem, and B. S. Kim, "Energy and Congestion-Aware Routing Metric for Smart Grid AMI Networks in Smart City," *IEEE Access*, pp. 13799–13810, 2017.
[22] M. Scarpiniti, "Fog of Everything : Energy-efficient Networked Computing Architectures , Research Challenges , and a Case Study Fog of Everything : energy-efficient networked computing architectures , research challenges , and a case study," no. May, 2017.
[23] K. Birman, L. Ganesh, and R. Renesse, "Running smart grid control software on cloud computing architectures," *Proc. Work. Comput. Needs Next Gener. Electr. Grid*, pp. 1–28, 2011.
[24] S. Bera, S. Misra, and D. Chatterjee, "C2C: Community-Based Cooperative Energy Consumption in Smart Grid," *IEEE Trans. Smart Grid*, vol. 3053, no. c, pp. 1–1, 2017.
[25] S. Asri and B. Pranggono, "Impact of Distributed Denial-of-Service Attack on Advanced Metering Infrastructure," *Wirel. Pers. Commun.*, vol. 83, no. 3, pp. 2211–2223, 2015.
[26] P. D. Diamantoulakis, V. M. Kapinas, and G. K. Karagiannidis, "Big Data Analytics for Dynamic Energy Management in Smart Grids," vol. 2, no. 3, pp. 94–101, 2015.
[27] Y. Yang, "Security architecture and key technologies for power cloud computing," *Proc. 2011 Int. Conf. Transp. Mech. Electr. Eng.*, pp. 1717–1720, 2011.
[28] D. Alahakoon and X. Yu, "Smart Electricity Meter Data Intelligence for Future Energy Systems: A Survey," *IEEE Trans. Ind. Informatics*, vol. 12, no. 1, pp. 425–436, 2016.